\documentclass[12pt]{article}
\usepackage{graphics,bezier}
\begin{document}
\markright{Faster-than-$c$ signals, special relativity, and causality\hfil}
\def\Universita{Universit\`a}
\title{\bf \LARGE
Faster-than-$c$ signals, \\
special relativity, \\
and causality}
\author{Stefano Liberati~$^*$, Sebastiano Sonego~$^\dagger$, and
Matt Visser~$^\ddagger$\\[2mm]
{\small \it
\thanks{liberati@physics.umd.edu; http://www2.physics.umd.edu/\~{}liberati}
Gravitation Theory Group -- Department of Physics,}\\
{\small \it University of Maryland, College Park, MD 20742-4111, USA.}\\[4mm]
{\small \it
\thanks{sebastiano.sonego@uniud.it}
{\Universita} di Udine, Via delle Scienze 208, 33100 Udine, Italy.}\\[4mm]
{\small \it
\thanks{visser@kiwi.wustl.edu; http://www.physics.wustl.edu/\~{}visser}
Physics Department, Washington University,}\\
{\small \it Saint Louis, Missouri 63130-4899, USA.}\\ }
\date{{\small 15 November 2001; {\LaTeX-ed \today}}}
\maketitle
\begin{abstract}
Motivated by the recent attention on superluminal phenomena, we
investigate the compatibility between faster-than-$c$ propagation and
the fundamental principles of relativity and causality.  We first
argue that special relativity can easily accommodate --- indeed, does
not exclude --- faster-than-$c$ signalling at the kinematical level.
As far as causality is concerned, it is impossible to make statements
of general validity, without specifying at least some features of the
tachyonic propagation. We thus focus on the Scharnhorst effect
(faster-than-$c$ photon propagation in the Casimir vacuum), which is
perhaps the most plausible candidate for a physically sound
realization of these phenomena.  We demonstrate that in this case the
faster-than-$c$ aspects are ``benign'' and constrained in such a
manner as to {\emph{not}} automatically lead to causality violations.

\vspace*{5mm} \noindent PACS: 03.30.+p; 04.20.Gz; 11.30.Cp; 11.55.Fv 
\end{abstract}
\vfill
\def\g{{\mbox{\sl g}}}
\def\Box{\nabla^2}
\def\d{{\mathrm d}}
\def\half{{1\over2}}
\def\quarter{{1\over4}}
\def\L{{\cal L}}
\def\sech{\hbox{sech}}
\def\SIZE{1.00}
\def\E{{\cal E}}
\def\ie{{\em i.e.\/}}
\def\eg{{\em e.g.\/}}
\def\etc{{\em etc.\/}}
\def\etal{{\em et al.\/}}
\def\aether{{{\ae}ther}}
\def\relative{{\mathrm{relative}}}
\def\Scharnhorst{{\mathrm{Scharnhorst}}}
\def\light{{\mathrm{light}}}
\def\gravity{{\mathrm{gravity}}}
\clearpage
\section{Introduction}
\setcounter{equation}{0}

Recently, there has been a renewed interest in issues related to
superluminal~\footnote{%
Hereafter, by ``superluminal'' we always mean ``faster than light in
unbounded empty space'', \ie, with speed larger than
$c=2.99792458\times 10^8\,\mbox{m s}^{-1}$. This terminology may
produce some weird-sounding sentences, such as ``light travels at a
superluminal speed in the Casimir vacuum'', but it cannot lead to any
misunderstanding if interpreted in the strictly technical sense
described above.}
propagation. This activity is connected with the experimental
realization of faster-than-$c$ group velocities~\cite{mugnai}, and
with the theoretical discovery of the Scharnhorst effect ---
faster-than-$c$ photon propagation in the Casimir vacuum due to higher
order QED corrections~\cite{scharnhorst,%
barton-scharnhorst,scharnhorst-bis,oblique}.  Although these two
strands of research deal with apparently similar subjects, their
implications on fundamental physics are quite different.  The
resonance-induced superluminal group velocities currently of
experimental interest were already known, theoretically, in the
sixties~\cite{brillouin}.  In spite of their large value (about
300 times $c$) and the many claims made in the non-specialized
press, they do not create any problem of principle, because the
{\em signal\/} velocity~\footnote{%
Following common usage, we identify the ``signal
velocity'' with the phase velocity at infinite frequencies,
equivalent to the so-called ``front velocity''
\cite{scharnhorst-bis}. See \cite{chiao} for other possible
definitions.}
still has $c$ as an upper
bound~\cite{fox1,bers,diener,recamietal,apparent-superluminal}.

In contrast, the Scharnhorst effect, which predicts that in a
cavity with perfectly reflecting boundaries, photons can travel at
a speed slightly larger than $c$, is an extremely tiny phenomenon,
well below our capabilities for experimental verification.  It is
nevertheless of fundamental theoretical importance, in that here
it seems to be the actual signal speed that it is modified.
Indeed, though the original derivations of the effect
\cite{scharnhorst} are carried out in the ``soft photon'' regime
(wavelengths much larger than the electron Compton wavelength),
there is an argument (based on the Kramers-Kronig dispersion
relations) which strongly suggests that the actual signal speed is
enhanced with respect to the ordinary speed of
light~\cite{barton-scharnhorst}.  In the present paper, we adopt
the view that these calculations, taken together, actually imply
that the Scharnhorst effect is physical, and not an artifact of
some approximation: it appears that quantum-polarization can
induce {\em true\/} superluminal velocities, albeit well outside
the realm of present day experimental techniques.

Regarding the generality of the effect, we stress that similar
results have been found for the propagation of photons in a
gravitational field~\cite{drummond-hathrell} (note that also these
calculations are based on the ``soft-photon'' approximation and
their extendibility to higher frequencies is still uncertain). A
framework which allows to discuss both types of effects in a
unified fashion has been recently developed
in~\cite{latorre,dittrich-gies}; see also~\cite{DGbook}.

The Scharnhorst effect, and its gravitational versions, are at first
disturbing because they indicate a violation of Lorentz invariance for
the electromagnetic field in vacuum, which is commonly taken as a
paradigmatic example of a Lorentz-invariant system.  Of course, in the
general relativistic framework of~\cite{drummond-hathrell} it is {\em
local\/} Lorentz invariance~\cite{will} that is violated.  Upon closer
inspection, however, Lorentz invariance is not affected at a
fundamental level.  Between conducting plates, or in a background
gravitational field, light does not propagate at the usual speed,
simply because the boundaries (or the background) single out a
preferred rest frame, which shows up in the property of the quantum
vacuum of not being Lorentz-invariant.  Thus, light behaves in a
non-Lorentz-invariant way only because the ground state of the
electromagnetic field is not Lorentz-invariant.  The Euler--Heisenberg
Lagrangian, from which the existence of the effect can be deduced, as
well as all the machinery of QED employed in its derivation, are still
fully Lorentz-invariant.  For this reason, one often speaks of a
``soft breaking'' of Lorentz invariance, in order to distinguish from
a situation in which also the basic equations, and not just the ground
state, are no longer Lorentz-invariant.

Of course, this soft breaking of Lorentz invariance has no
fundamental influence on special relativity --- no more than being
inside a material medium has.  For many purposes, the quantum
vacuum can indeed be regarded as a medium --- an
``\aether''~\cite{dewitt}. The fact that the Scharnhorst effect
appears to violate the principle of relativity can then be
understood by considering that, between infinite parallel plates,
the {\aether} is no longer Lorentz-invariant, so its presence, in
contrast to the situation in infinite space with no boundaries,
{\em can\/} be detected.  Nevertheless one can always imagine
experiments based on interactions which are unaffected by the QED
vacuum, such as gravity, that would show no violation of Lorentz
invariance. Only if the deviations were universal, which is not
the case, would special relativity be threatened (and even in that
case, one could easily avoid troubles by adopting a metric
different from the Minkowski one, \ie, by shifting to a {\em
general\/} relativistic context).

There is another unpleasant feature of Scharnhorst's result,
though. As we pointed out, any refractive medium leads to a soft
breaking of Lorentz invariance, as shown by the fact that the speed of
light differs from the value $c$.  However, in all other known cases,
the speed of light is {\em smaller\/} than $c$, while, according to
Scharnhorst, between plates it should be {\em greater\/}.  This is a
conceptually nontrivial result, because it contradicts the common
belief that signal propagation at speeds larger than the speed of
light in vacuum is forbidden by special relativity.  Furthermore, it
shows that faster-than-$c$ effects can occur even within
well-established theories such as QED.

The main goal of the present paper is to show that faster-than-$c$
propagation does not violate the basic tenets of special relativity,
and is kinematically perfectly compatible with it
(section~\ref{S:sr}). In addition, we shall prove in
section~\ref{S:causality} that the existence of phenomena such as the
Scharnhorst effect does not necessarily lead to closed causal curves,
contrary to some claims made in the literature~\cite{dolgov-novikov}.
In particular, for a single pair of Casimir plates photon propagation
can be described by an ``effective metric'' that exhibits the property
of ``stable causality'', which automatically prevents the formation of
closed timelike loops. When one considers multiple pairs of Casimir
plates, naive attempts at producing causality violations are vitiated
by their need for either grossly unphysical inter-penetrating plates
(which undermine the whole basis for the Scharnhorst calculation), or
by violent and uncontrolled edge effects. The general situation is
best analyzed in terms of Hawking's chronology protection conjecture.

Section~\ref{S:conclusion} contains a brief summary of the main
results of the paper.  Although it does not challenge present-day
physics, the possibility of faster-than-$c$ signalling
nevertheless brings out some interesting open issues about the
foundations of special relativity.  Since these issues, although
related to some points touched in our discussion, are not directly
relevant for the main topic of the paper, we present them in the
three appendices.

\section{Special relativity}
\setcounter{equation}{0}
\label{S:sr}

There is a logical distinction between the concept of an {\em
invariant\/} speed, and that of a {\em maximum\/} speed. Contrary
to widespread belief, special relativity only requires, for its
kinematical consistency, that there be an invariant speed. This is
already obvious even from the original derivation by Einstein of
the Lorentz transformations~\cite{einstein}. However, Einstein's
derivation relies upon a procedure of clock synchronization in
which light signals are used, and one might think that if
faster-than-light propagation is possible, it could be used to
synchronize clocks in an alternative way, thus undermining
relativistic kinematics at its very foundations.  In this section
we show that this is not the case.

\subsection{Lorentz transformations without light}
\label{Ss:ignat}

We find it convenient to start from an alternative, less
well-known derivation of the Lorentz transformations, first
discussed by von Ignatowsky in 1910~\cite{ignat1}, and later
rediscovered by many people~\cite{fr1,terletskii,sussmann,%
berzi,zecca,leekalotas,ll,rindler,jammer,torretti}. The starting
hypotheses are:
\begin{description}
\item (i) That in an inertial frame space is homogeneous, isotropic,
and Euclidean, \\
and time is homogeneous;
\item (ii) The principle of relativity;
\item (iii) A condition of ``pre-causality'' (see below).
\end{description}
These postulates alone imply the existence of the Lorentz group,
containing {\emph{some}} parameter $C>0$ that represents an {\em
invariant\/} speed. We now sketch the main lines of the
argument.  A rigorous and detailed presentation can be found in
the papers by Gorini and collaborators~\cite{berzi,zecca}; see
also the book by Torretti~\cite{torretti} for a good summary,
and~\cite{leekalotas,ll,rindler} for pedagogical introductions.

\begin{figure}%
\begin{picture}(100,250)(10,20)%
\put(105,90){$O$}%
\put(120,90){\vector(1,0){300}}%
\put(415,80){$x$}%
\put(120,90){\vector(0,1){130}}%
\put(105,215){$y$}%
\put(120,90){\vector(-1,-1){70}}%
\put(40,25){$z$}%
\put(205,95){$O'$}%
\put(220,95){\vector(1,0){200}}%
\put(415,100){$x'$}%
\put(220,95){\vector(0,1){125}}%
\put(205,215){$y'$}%
\put(220,95){\vector(-1,-1){72}}%
\put(135,25){$z'$}%
\put(123,120){\vector(1,0){95}}%
\put(165,125){$vt$}%
\put(222,85){\vector(1,0){150}}%
\put(300,70){$\xi$}%
\end{picture}%
\caption{\small Spatial distances as measured in the reference frame
$\cal K$; see equation~(\ref{xi}).}%
\label{fig1}%
\end{figure}
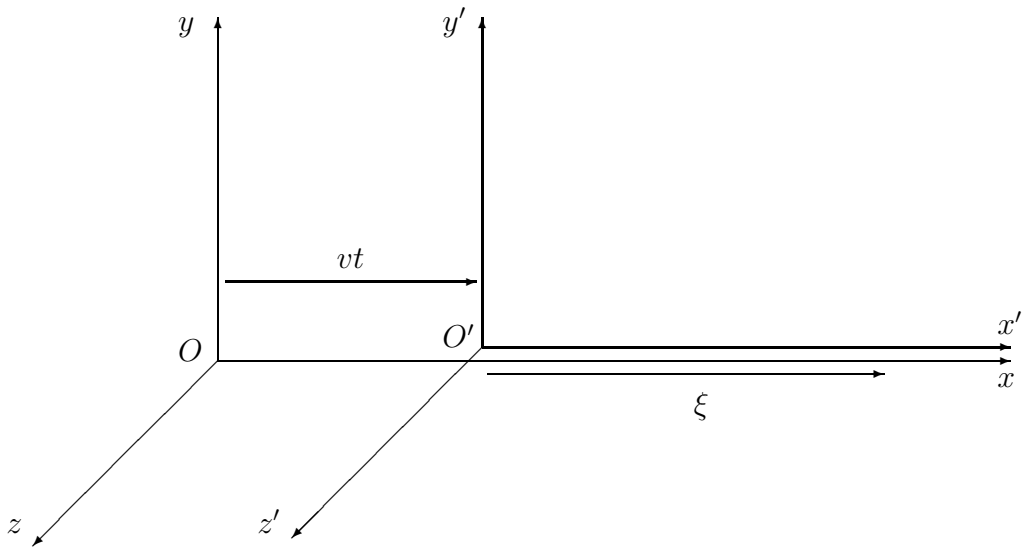%

Let us suppose that, in two inertial frames $\cal K$ and ${\cal
K}'$, with spatial origins $O$ and $O'$, respectively, the same
event is labelled by coordinates $(t,x,y,z)$ and $(t',x',y',z')$.
Because of isotropy of space, we can orient the spatial axes so
that the relative velocity between the frames is along $x$ and
$x'$. Since the relationship between the other coordinates turns
out to be trivial, we shall ignore them, assuming that
$y=y'=z=z'=0$, and focus only upon $(t,x)$ and $(t',x')$.

According to the reference frame $\cal K$, the spatial distance
between the event and the origin $O$ is obtained by summing the
distance of the event from $O'$ measured in $\cal K$, say $\xi$, to
the distance between $O$ and $O'$.  The latter is simply $vt$, where
$v$ is the velocity of $O'$ with respect to $\cal K$, so
\begin{equation}%
x=\xi+vt%
\label{xi}\end{equation}%
(see figure~\ref{fig1}).  Similarly, if we call $\xi'$ the
distance between the event and $O$ according to ${\cal K}'$, and
$v'$ the velocity of $O$ with respect to ${\cal K}'$, we have
\begin{equation}%
x'=\xi'+v't'\;.%
\label{xi'}\end{equation}%
Now, homogeneity of space requires that the relationship between the
distance of the event from $O'$, measured in the two frames, be
linear.  Thus, there exists a constant coefficient $\gamma$ (possibly
dependent on $v$ and $v'$) such that $\xi=x'/\gamma$.  Similarly,
there will be a $\gamma'$ such that $\xi'=x/\gamma'$.  Of course,
$\gamma$ and $\gamma'$ must be positive, because they relate
measurements of distances.  Replacing these expressions into
equations~(\ref{xi}) and (\ref{xi'}), and solving with respect to $t'$
and $x'$, we get:
\begin{eqnarray}%
t'&=&-{\gamma v\over v'}\left(t-{\gamma\gamma'-
1\over\gamma\gamma'v}\,x\right)\;;%
\label{t'}
\\
x'&=&\gamma\left(x-vt\right)\;.%
\label{x'}
\end{eqnarray}%

Until now, we have only used the hypothesis of homogeneity and
isotropy of space.  A second ingredient in the derivation is the
so-called {\em reciprocity principle\/}, which asserts that
$v'=-v$. (This is a non-trivial consequence of the isotropy of
space~\cite{berzi,ll,torretti}.) In addition, the principle of
relativity imposes the further constraint that $\gamma'=\gamma$,
and it requires that the transformations form a group.  {From} the
latter condition one can then easily show, by considering three
inertial frames, that $\gamma^{-2}=1-kv^2$, with $k$ a
$v$-independent constant.

The case of a negative $k$ can be excluded by the following
argument.  If $k<0$, on defining $\tau:=t/\sqrt{-k}$ and
$\tan^2\psi:=-kv^2$, the resulting transformation is easily
recognized to be an ordinary rotation of an angle $\psi$ in the
$(\tau,x)$ plane. This has a number of unphysical consequences. In
particular, let us consider two events with the same $x$
coordinate in $\cal K$, and different $t$ coordinates; then it is
always possible to change reference frame in such a way that the
time order of the
events is reversed.\footnote{%
This is most easily seen by applying two successive
transformations with ``speeds'' $v_1$ and $v_2$, say, such that
$v_1v_2>-1/k$.}
This result would clearly prevent the application of any meaningful
notion of causality.  In order to avoid this possibility, we impose
the following ``pre-causality'' requirement, which is a necessary
condition for causality to be defined: {\em If two events happen at
the same place in one reference frame, their time order must be the
same in all reference frames}.\footnote{%
Equivalently, one might ask that the composition of two velocities
in the same direction does not produce a velocity pointing in the
opposite direction~\cite{ll,rindler}.}
This implies $k\geq 0$, so we can define a (possibly infinite)
positive constant $C$ such that $k=1/C^2$. The transformation
(\ref{t'})--(\ref{x'}) thus takes on the familiar Lorentz form,
with the speed of light replaced by the constant $C$.  It is then
obvious, considering the associated composition law for
velocities, that if a signal travels at the speed $C$ in one
reference frame, it does so in all reference frames. That is, $C$
is an invariant speed.

Note that nowhere in the previous argument have we assumed the
existence of a maximum speed. The principle of relativity,
together with the requirements of homogeneity and isotropy, is
sufficient in order to obtain a Lorentz-like transformation. Nor
have we deduced that signals cannot propagate at a speed greater
than $C$. We have only found that one cannot consider {\em
reference frames\/} in motion with relative speed $v>C$ (otherwise
$\gamma$ would turn out to be imaginary), but of course this does
not mean that there is an upper bound to signal velocity. Even if
it were possible to send signals at arbitrarily high speeds, the
kinematical group could still have the Lorentz form with a finite
value of $C$. This is a counterintuitive result, because one might
think that if arbitrarily fast signalling were possible, then one
should be able to synchronize clocks in an absolute way.

\subsection{Fluid-dynamical analog}

Let us try to clarify this point by using an analogy. Consider a fluid
at rest with respect to an inertial frame, within the context of {\em
Newtonian\/} mechanics, and two reference frames, one coinciding with
the rest frame of the fluid, the other moving with respect to it with
a constant velocity ${\bf v}$. Furthermore, let us assume that the
axes of the two frames are parallel, and that ${\bf v}$ is directed
along one of them, so we don't need to care about the other two. If
the observers in the two frames agree to use ultra-fast signals
(certainly allowed by Newtonian physics) in order to synchronize their
clocks, the relationship between the coordinates $(T,X)$ of the frame
at rest with respect to the fluid, and the coordinates $(t_G,x_G)$ of
the other frame is the usual Galilean transformation:
\begin{eqnarray}%
&& t_G=T\;;\label{tG}\\%
&& x_G=X-vT\;.\label{xG}%
\end{eqnarray}%
Using these coordinates, the propagation of waves in the fluid
will be described by the two frames in a different way.  For two
events on the wavefront of a sound wave propagating along the $X$
direction, separated by coordinate lapses $\Delta T$ and $\Delta
X$, we have
\begin{equation}%
-c_s^2 \Delta T^2+\Delta X^2=0\;,%
\label{wavefront}\end{equation}%
where $c_s$ is the speed of sound in the frame at rest with
respect to the fluid.  On the other hand, in $(t_G,x_G)$
coordinates we have
\begin{equation}%
-c_s^2\Delta t_G^2+ \left(\Delta x_G+
v\Delta t_G\right)^2=0\;,%
\label{wavefrontG}\end{equation}%
where we have inserted the Galilean transformations (\ref{tG}) and
(\ref{xG}) into equation~(\ref{wavefront}). Obviously,
equation~(\ref{wavefrontG}) does not have the same form as
equation~(\ref{wavefront}), even replacing $c_s$ by some new function
of $c_s$ and $v$.  Thus, although the Galilean transformation does
satisfy the principle of relativity, the description of sound
propagation given by two sets of coordinates linked by a Galilean
transformation is not symmetric.  (Of course, there is no reason why
it should be, because the presence of the fluid causes a soft breaking
of Galilean invariance, since it defines an ``absolute'' rest frame.)

However, the moving observer could decide to {\em define\/} his
coordinates in such a way that the principle of relativity is still
satisfied {\em and\/} that the invariant speed is no longer
$C=+\infty$, as in the Galilean transformation, but $C=c_s$. To do
this, it is sufficient that he use sound signals in order to
synchronize his clocks, and to define distances by an acoustic ``radar
procedure'', as bats would do (this is possible if $v<c_s$). Calling
$(t_L,x_L)$ the coordinates defined in this way, we have
\begin{eqnarray}%
&& t_L=\gamma_s\left(T-vX/c_s^2\right)\;,\label{tL}\\%
&& x_L=\gamma_s\left(X-vT\right)\;,\label{xL}%
\end{eqnarray}%
where $\gamma_s=\left(1-v^2/c_s^2\right)^{-1/2}$.  Thus, the
coordinates $(t_L,x_L)$ are related to $(T,X)$ by a Lorentz
transformation with invariant speed $c_s$.  It is obvious that the
relationship between the lapses $\Delta t_L$ and $\Delta x_L$
corresponding to two events on a sound wavefront is now
\begin{equation}%
-c_s^2 \Delta t_L^2+\Delta x_L^2=0\;,%
\label{wavefrontL}\end{equation}%
\ie, it has exactly the same form as
equation~(\ref{wavefront}).\footnote{%
As a side remark about this analog, notice that one could introduce a
matrix with coefficients $G_{\mu\nu}$ such that $G_{\mu\nu}\; \Delta
x^\mu\; \Delta x^\nu=-c_s^2 \Delta T^2+\Delta X^2=0$ in {\em any\/}
system of coordinates $x^\mu$. One then gets, in Galilean coordinates,
\begin{equation}%
\left[G_{\mu\nu}\right]=\left[
\begin{array}{cc}
-\left(c_s^2-v^2\right) & v \\
v & 1
\end{array}
\right]\;,
\end{equation}%
which is just the so-called acoustic metric, describing the
propagation of perturbations in a fluid that moves with velocity
$-{\bf v}$~\cite{acoustic}.}

This example brings out a fundamental issue: Given a propagation
phenomenon that is associated with an equation like
(\ref{wavefront}) in one reference frame, one can {\em always\/}
find time and space coordinates, in any other frame, such that the
principle of relativity is satisfied and an analogous equation
holds true in the new frame.  Thus, the Lorentz transformations
corresponding to an arbitrary invariant speed $C$ are, in general,
devoid of physical meaning.  One can always {\em choose\/} the
value of $C$, because such a choice is essentially equivalent to
giving a prescription for synchronization, selecting the class of
signals used in order to synchronize clocks.\footnote{%
In this respect, note that in equations~(\ref{tL}) and (\ref{xL}) the
factor $\gamma_s$ re-defines units, while the term $vX/c_s^2$
corresponds to a choice of synchronization.}
One is then led to ask: Is not the situation just the same with the
Lorentz transformations? If so, is the invariance of the speed of
light just a convention, with no physical content?

\subsection{Physical coordinates}
\label{physcoord}

The missing link is given by the {\emph{interpretation}} of the time
and space coordinates. In general, one establishes a reference frame
first by choosing physical units of time and distance, based on some
physical phenomenon. Then, a Cartesian lattice is constructed using
the standards of length, and clocks are (ideally) placed at all points
in space.  All these clocks tick at the same rate, but they are not
yet synchronized.  However, it is already possible to decide
empirically whether some signals have an invariant speed.  It is
sufficient that an observer sitting at a point $P$, say, send a signal
to another observer at a point $Q$ of the same frame, and that this
one immediately signal back to $P$.  On measuring the time taken by
the round trip on his clock, and knowing his distance from $Q$, the
observer at $P$ can readily establish the average value of the signal
speed.\footnote{%
Note that the one-way speed cannot be measured unless clocks at
different places are synchronized~\cite{jammer,reichenbach,%
sexl,erlichson,brehme1,ungar,anderson}.}
If, repeating this kind of measurement along different directions, and
in different frames equipped with the same type of clocks and rods,
one always obtains the same value, these signals travel at the
invariant speed.  In the Newtonian fluid dynamic example of the
previous subsection, this speed is infinite.  It is important to
emphasize that, since units of time and distance have been chosen in
advance, the value of the invariant speed cannot be chosen at one's
will.  Hence, the existence of an invariant speed of some specific
value, albeit a two-way average speed, is a nontrivial physical law,
and not a consequence of an arbitrary convention (see
Reichenbach~\cite{reichenbach}, pp.~204--205).

The next step is the prescription for the synchronization of
clocks at different places --- hence, for the definition of
simultaneous events in a given reference frame.  This {\em is\/} a
conventional element in the theory~\cite{jammer,reichenbach,%
sexl,erlichson,ungar,anderson} (however, see \cite{brehme1} for a
different opinion).  It is clearly embodied in the postulates (i)
and (ii), which imply a well-defined relationship between the
coordinates in two different inertial frames, and hence a
well-defined notion of which events are simultaneous in an
arbitrary frame.  This, in turn, is equivalent to postulating
that the speed $C$ be invariant not only on average during a
round trip, but also in the one-way sense.

Of course, one could use the signals travelling at the invariant
speed in order to synchronize clocks, but we want to understand
what would happen using other signals. It is easy to see, however,
that if the coordinates in the two reference frames are related by
a Lorentz transformation with invariant speed $C$, synchronization
{\em must\/} be performed using signals that travel at the speed
$C$. In order to prove this, let us consider again the framework
described in section~\ref{Ss:ignat} (see figure~\ref{fig1}), and
let us suppose that both reference frames agree to use signals
that travel at some given speed $V$ in $\cal K$. Then, if an
observer at $O'$ sends two such signals along the axis $x'$ in
opposite directions, they will reach observers in ${\cal K}'$ with
$x'$ coordinate equal to $l$ and $-l$, say, at the times
\begin{equation}%
t'_+=l\,{1-Vv/C^2\over V-v}%
\end{equation}%
and
\begin{equation}%
t'_-=l\,{1+Vv/C^2\over V+v}\;,%
\end{equation}%
respectively.  (Here, we have applied the law of composition of
velocities that follows by the assumption that the coordinates in
$\cal K$ and ${\cal K}'$ are related by a Lorentz transformation with
invariant speed $C$.) If signals travelling at speed $V$ have been
used to synchronize clocks, then we must have $t'_+=t'_-$, which is
possible only if $V=C$.

Now there are two possibilities.  If the value of the invariant
speed has already been established experimentally, using the
method outlined above, this result forces us to use precisely the
invariant signals, and not others, in order to synchronize clocks.
Alternatively, we might want $V$ to be the invariant speed, as in
the example of the fluid, where we defined $(t_L,x_L)$ in such a
way that $c_s$ turns out to be invariant. Of course this is
possible, but it requires the use of time and space coordinates
that behave differently from those associated with the physical
clocks and rods.\footnote{%
This can be codified into a somewhat unusual ``take'' on special
relativity: The Maxwell equations, considered simply as a
mathematical system, possess a symmetry, the Lorentz group, under
redefinitions of the labels $x$ and $t$. But this is a purely
mathematical statement devoid of interesting physical consequences
until one asks how physical clocks and rulers are constructed, and
what forces hold them together. Since it is electromagnetic forces
balanced against quantum physics which holds the internal
structure of these objects together, the experimental observation
that to very high accuracy physical bodies also exhibit Lorentz
symmetry allows one to deduce that quantum physics obeys the same
symmetry as the Maxwell equations. Viewed in this way, all
experimental tests of special relativity are really precision
experimental tests of the symmetry group of quantum physics.}
In other words, while those Lorentz transformations associated with
the physical invariant speed measured using independent standards of
time and length connect actual measurements performed in two
different frames, all other logically possible Lorentz transformations
must be regarded as mere changes in the labels for the events, with no
physical meaning.  For example, in our fluid analog, the physical
measurements of time and space intervals are $\Delta T$ and $\Delta X$
in all reference frames, according to the postulates of Newtonian
mechanics.  This selects automatically the transformation with
$C=+\infty$ as the physically relevant one, among all those allowed by
the relativity principle alone.

However, it is interesting to notice that if the observers were
restricted to using the fluid in their measurements, {\em and
nothing else\/}, the ``physical'' coordinates would probably turn
out to be $t_L$ and $x_L$.  Indeed, hypothetical beings
whose internal structure was completely mediated by phonon
exchange, and whose rulers and clocks were likewise held together
by phonon exchange, and who were completely blind to
electromagnetism (so in particular they could not probe the atomic
structure of individual atoms) would discover an ``acoustic
Lorentz invariance'' with their ``rulers'' and ``clocks''
transforming according to the laws of an approximate ``acoustic
relativity'' (the Lorentz group with $C=c_s$).\footnote{%
Of course for normal observers, made of normal condensed matter,
physical rods and clocks are seen (to good approximation) to
transform according to the $C=c$ Lorentz group; normal observers
would not see this ``acoustic relativity''.}
In this case one might say that the measuring devices are
dynamically affected by the motion with respect to the fluid, in
such a way that such motion becomes undetectable.  In a sense, the
fluid would play the role of an {\aether} whose presence is masked
by length contraction and time dilation effects.  Thus, we would
have a fluid dynamical analog of special relativity {\em \`a la\/}
Lorentz~\cite{lorentz}.\footnote{%
Note that such an analogy cannot be strictly true in the case of
the fluid. As a matter of fact, measurements performed by our
hypothetical ``fluid observers'' would be insensitive to
non-hydrodynamical physics only if they were probing a regime
where the symmetry group of quantum physics (presumably the
Lorentz group with $C=c$) did not overwhelm the acoustic effects.
Moreover, even remaining in a classical regime, fluids definitely
possess a minimum length scale (of the order of the intermolecular
distance) under which the hydrodynamical approximation breaks
down. This fact would by itself lead to a violation of Lorentz
invariance at small scales and modifications of the dispersion
relation for phonons at high frequencies. We shall come back to
this point in our appendix A.}

\subsection{What is $C$?}

Let us summarize.  Homogeneity and isotropy of space in an inertial
frame, together with the principles of relativity and causality, imply
the kinematical Lorentz group, hence the existence of some invariant
speed $C$.  Clocks at a distance must be synchronized using signals
that propagate at the speed $C$.  The value of $C$ is fixed once a
prescription for constructing clocks and rulers is given and adopted
uniformly in all reference frames.

Since the argument that establishes the existence of an invariant
speed $C$ is entirely of a kinematical character, it does not
allow one to know the value of $C$. In principle, one could even
have the degenerate case $C=+\infty$, which corresponds to
Galilean relativity.\footnote{%
It has been argued, within the context of constructive axiomatic
theories of spacetime structure, that signals with a maximum speed
must exist, if one wishes to introduce coordinates by a radar
technique~\cite{schelb}.  Of course, this indicates a difficulty of
the radar technique in the case $C=+\infty$, rather than the logical
impossibility of signals with infinite speed.  Newtonian spacetime
provides an example of a consistent spacetime model to which the
axiomatic framework of reference~\cite{schelb} cannot be applied.
Another problem for theories of this type arises precisely in the
Casimir vacuum. Since the maximum speed is used in order to define the
causal structure of spacetime, the existence of the Scharnhorst effect
should lead one to adopt, inside plates, a spacetime metric different
from the one outside.  Thus, one would have a modification of the
spacetime structure in a situation where the gravitational effects are
certainly negligible.}
To determine the actual value of $C$ when a choice for clocks and
rulers has been made is an experimental problem. We know that $C$
coincides, to a very good accuracy, with the value $c$ of the speed of
light in a vacuum with no boundaries, so we shall set $C\equiv c$ in
the rest of this paper. However, one should always keep in mind that
tiny deviations from this empirically established equality are
logically possible.

Once a finite value for $C$ is found experimentally, the whole
formalism of special relativity follows on the basis of postulates
(i)--(iii).  Since faster-than-$C$ signals do not contradict any of
these postulates, their possible existence is (kinematically)
perfectly acceptable within ordinary special relativity.  In
particular, one could not make use of these signals in order to
synchronize clocks without violating (i)--(ii) above.  In fact,
although different frames {\em could\/} use hypothetical ultra-fast
signals in order to define the time coordinates $t$ and $t'$ in such a
way that $\Delta t=0$ whenever $\Delta t'=0$, the use of such
coordinates would be incompatible with the principle of relativity.
Indeed, if $\Delta t=\Delta t'=0$, equation~(\ref{t'}) implies
$\gamma\gamma'=1$ and $t'=\left(-\gamma v/v'\right)t$.  Assuming
isotropy of space gives the reciprocity relation $v'=-v$, so
$t'=\gamma t$. If $\gamma\neq 1$ --- an experimental issue that can be
decided by comparing coordinate clocks in the two frames --- this
requires $\gamma'\neq\gamma$, in violation of the principle of
relativity.  The resulting transformation allows, of course, for an
absolute notion of simultaneity, essentially due to the introduction
of a preferred frame in the
formalism~\cite{sexl,anderson,tangherlini}.  Note that if $\gamma\neq
1$ it is impossible to set clocks so that $t=t'$, unless one accepts
that reciprocity, hence isotropy of space, is also violated.

Hence, even if there were tachyons (\ie, particles or signals
faster than $C$), one could not use them in order to synchronize
clocks in a frame-independent way, without contradicting the
relativity principle. Of course, they could be used to define some
$t$ and $x$ coordinates --- just as {\em any\/} other signal
(see~\cite{rindler}, p.~30). But for synchronization one must use
signals travelling at the invariant speed $C$, if one wants to
satisfy the principle of relativity.

\section{Causality}
\setcounter{equation}{0}
\label{S:causality}

We have seen in the previous section that, once units of length
and time are defined, purely kinematical arguments based on
postulates (i)--(iii) lead to the existence of an invariant speed.
Experimentally, the latter turns out to coincide with the
speed of light $c$ when we use ``ordinary'' clocks and rods. This,
however, does not necessarily mean that nothing can propagate
faster than $c$. As we have already discussed, the existence of
signals faster than $c$ would not threaten kinematic features of
special relativity, because they could not be used to synchronize
clocks in a physically acceptable way.  Hence, a model of the
world based on a four-dimensional spacetime with the Minkowski
metric
\begin{equation}%
\eta=-c^2\d t^2+\d x^2+\d y^2+\d z^2\;,%
\label{mink}\end{equation}%
in which particles can travel along timelike, null, and spacelike
lines, is perfectly consistent from a kinematical point of
view~\cite{terletskii,camenzind}.  In other words, special
relativity can kinematically accommodate faster-than-$c$
propagation.\footnote{%
It is nevertheless worth noticing that, although faster-than-$c$
signals are not incompatible with special relativity, there is
apparently a structural obstruction to forming stable
tachyons~\cite{bers}.}

\subsection{Tachyons and causal paradoxes}

On the other hand, tachyons are usually associated with unpleasant
causal paradoxes.  The basic reason for this belief is that if two
events, say ${\cal E}_1$ and ${\cal E}_2$, are spacelike related,
there is no absolute time ordering between them. Thus, if a signal
travels from ${\cal E}_1$ to ${\cal E}_2$ in a reference frame, it is
always possible to find another frame where ${\cal E}_1$ and ${\cal
E}_2$ are simultaneous, so the signal would appear to travel at an
infinite speed, and others where ${\cal E}_1$ happens after ${\cal
E}_2$, so the signal would ``travel to the past''.\footnote{%
We are of course currently using the term ``tachyon'' in the sense of
some postulated faster-than-$c$ particle or signal. There is a subtly
different usage of the word ``tachyon'' in a field-theoretic sense to
refer to a field theory with canonical kinetic energy and negative
mass-squared (imaginary proper mass). Despite the fact that the
resulting dispersion relation mimics the mass-momentum relation of a
faster-than-$c$ particle, field-theoretic tachyons of this particular
type cannot be used to send faster-than-$c$ signals. Classically this
arises because the characteristics of the field equations (the partial
differential equations derived from the field theory) depend only on
the kinetic energy term and are completely insensitive to the mass
term. As long as the kinetic energy terms are canonical the
characteristics are the usual ones, and the signal speed is $c$,
regardless of whether the proper mass is positive, zero, or
imaginary. This observation can be rephrased in terms of the evolution
of initial data of compact support, for which it is a standard result
of the theory of partial differential equations that canonical kinetic
energy terms limit the support of the evolved field configuration to
lie inside the light-cone of the initial data~\cite{bers}. This result
also holds quantum mechanically where canonical quantization requires
equal-time commutators to vanish, which then extends via Lorentz
invariance to vanishing of the field commutators everywhere outside
the light cone; thus quantum mechanically the signal speed is also
limited to $c$. For field-theoretic ``tachyons'' with canonical
kinetic energies it is not causality that is the problem, rather it is
the issue of the instability of the field-theory ground state that
leads to difficulties. Note that for the Scharnhorst photons we are
chiefly concerned with in this article, one-loop quantum effects modify
the kinetic energy terms so that they are not canonical; this shifts
the characteristic surfaces, and so shifts the signal speed so that it
is no longer equal to $c$.}

Obviously, this argument could be used to criticize faster-than-$c$
propagation only if, in any given reference frame, the only criterion
for saying that an event is the cause of another one were the time
ordering in that frame.  However, cause and effect are usually {\em
not\/} defined in this way.\footnote{%
Note that, if a causal link between two events could only be
established on the basis of their temporal order, the notions of cause
and effect would be problematic in theories (such as, \eg, Newtonian
gravity) which admit instantaneous action at a distance.}
In fact, there are no precise definitions of these concepts, but only
some intuitive ideas that allow us to recognize, in some cases, the
existence of a causal relationship between events.  Eventually, the
criteria used to establish that ${\cal E}_1$ is a cause of ${\cal
E}_2$ are based on considerations of complexity of the type usually
involved in discussions about the so-called ``arrow of time''.  For
example, if ${\cal E}_1$ represents the emission of a signal from a
broadcasting station, and ${\cal E}_2$ its reception on a TV set, we
consider ${\cal E}_1$ a cause of ${\cal E}_2$ not just because it
happened at an earlier time, but mainly because the opposite choice
would require the presence of weird, conspiratorial, correlations.
Now, if ${\cal E}_1$ and ${\cal E}_2$ were connected by
faster-than-$c$ photons instead of ordinary ones, the same criterion
would apply, so there would be an absolute notion of what is cause and
what is effect.  In this respect, the exchange in the time ordering of
${\cal E}_1$ and ${\cal E}_2$ within some reference frames should not
be more disturbing than the jet lag experienced by travellers because
of the peculiar way clocks are set on the Earth.\footnote{%
The only case in which an inversion of the time ordering could
really be problematic for causality is when the events happen at
the same place in one reference frame.  This possibility is
excluded for transformations that fulfill the pre-causality
condition of section~\ref{Ss:ignat}, including of course the
Lorentz ones.}
Note, however, that if ${\cal E}_1$ causes ${\cal E}_2$, there is
always at least one frame in which ${\cal E}_2$ does not happen
earlier than ${\cal E}_1$.

The inversion of the time ordering for two events connected by the
propagation of a tachyon is therefore not, by itself, a
difficulty. However, unless some restriction is imposed on the
type of propagation, it is potentially a source of paradoxes, as
it can lead to situations where two events are {\em timelike\/}
related, and yet the cause follows the effect.  A typical
argument is the  following, sometimes picturesquely referred
to as the ``tachyonic anti-telephone'' \cite{tolman}.

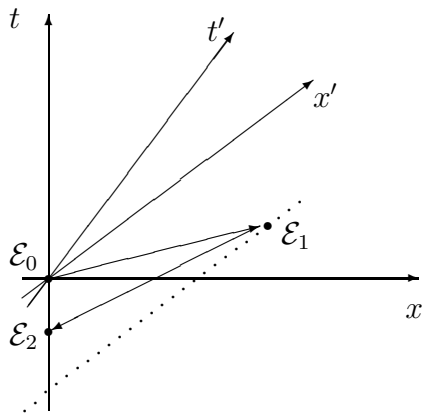
\begin{figure}%
\begin{picture}(150,200)(-30,50)%
\put(110,120){\vector(1,0){150}}%
\put(255,105){$x$}%
\put(120,70){\vector(0,1){150}}%
\put(105,215){$t$}%
\put(120,120){\circle*{3}}%
\put(105,125){${\cal E}_0$}%
\put(120,120){\vector(4,3){100}}%
\put(120,120){\line(-4,-3){10}}%
\put(220,185){$x'$}%
\put(120,120){\vector(3,4){70}}%
\put(120,120){\line(-3,-4){8}}%
\put(180,210){$t'$}%
\put(120,120){\vector(4,1){80}}%
\put(203,140){\circle*{3}}%
\put(208,133){${\cal E}_1$}%
\put(200,140){\vector(-2,-1){79}}%
\put(120,100){\circle*{3}}%
\put(105,95){${\cal E}_2$}%
\put(111,70){\circle*{1}}%
\put(115,73){\circle*{1}}%
\put(119,77){\circle*{1}}%
\put(123,80){\circle*{1}}%
\put(127,83){\circle*{1}}%
\put(131,86){\circle*{1}}%
\put(135,89){\circle*{1}}%
\put(139,92){\circle*{1}}%
\put(143,95){\circle*{1}}%
\put(147,98){\circle*{1}}%
\put(151,101){\circle*{1}}%
\put(155,104){\circle*{1}}%
\put(159,107){\circle*{1}}%
\put(163,110){\circle*{1}}%
\put(167,113){\circle*{1}}%
\put(171,116){\circle*{1}}%
\put(175,119){\circle*{1}}%
\put(179,122){\circle*{1}}%
\put(183,125){\circle*{1}}%
\put(187,128){\circle*{1}}%
\put(191,131){\circle*{1}}%
\put(195,134){\circle*{1}}%
\put(199,137){\circle*{1}}%
\put(203,140){\circle*{1}}%
\put(207,143){\circle*{1}}%
\put(211,146){\circle*{1}}%
\put(215,149){\circle*{1}}%
\end{picture}%
\caption{\small A causal paradox using tachyons. The dotted line
represents the set of events which are simultaneous with ${\cal
E}_1$ according to the reference frame ${\cal K}'$.  The tachyonic
signal from ${\cal E}_1$ to ${\cal E}_2$ travels to the future
with respect to ${\cal K}'$, and to the past with respect to
$\cal K$.}%
\label{fig2}%
\end{figure}%

Suppose that, in an inertial frame $\cal K$, a tachyon is emitted
at $t_0=0$, $x_0=0$ (event ${\cal E}_0$ in figure~\ref{fig2}), and
received at an event ${\cal E}_1$ with $t_1>0$.  It is always
possible to find another inertial frame ${\cal K}'$, in the
configuration considered in section~\ref{Ss:ignat}, such that
$t'_0=0$, $x'_0=0$, and $t'_1<0$.  Now, suppose that at the event
${\cal E}_1$ a second tachyon, that travels to the future with
respect to ${\cal K}'$ and to the past with respect to $\cal K$,
is sent toward the origin.  This reaches the spatial origin of
$\cal K$ (event ${\cal E}_2$) at a time $t_2<0$.  We can arrange
the experiments in such a way that ${\cal E}_0$ causes ${\cal
E}_1$ which, in turn, causes ${\cal E}_2$.  Therefore, ${\cal
E}_0$ causes ${\cal E}_2$, which is a paradoxical result because,
since these two events are timelike related and $t_0>t_2$, ${\cal
E}_0$ follows ${\cal E}_2$ in {\em all\/} reference frames.  More
elaborate versions of this paradox, based on the particular use of
Scharnhorst photons, have been presented in~\cite{dolgov-novikov}.

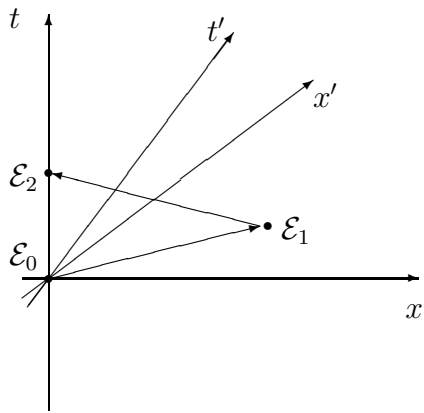
\begin{figure}%
\begin{picture}(150,200)(-30,50)%
\put(110,120){\vector(1,0){150}}%
\put(255,105){$x$}%
\put(120,70){\vector(0,1){150}}%
\put(105,215){$t$}%
\put(120,120){\circle*{3}}%
\put(105,125){${\cal E}_0$}%
\put(120,120){\vector(4,3){100}}%
\put(120,120){\line(-4,-3){10}}%
\put(220,185){$x'$}%
\put(120,120){\vector(3,4){70}}%
\put(120,120){\line(-3,-4){8}}%
\put(180,210){$t'$}%
\put(120,120){\vector(4,1){80}}%
\put(203,140){\circle*{3}}%
\put(208,135){${\cal E}_1$}%
\put(200,140){\vector(-4,1){79}}%
\put(120,160){\circle*{3}}%
\put(105,155){${\cal E}_2$}%
\end{picture}%
\caption{\small Tachyon propagation without causal paradoxes.  Both
signals travel to the future in at least one reference frame.}

\label{fig3}%
\end{figure}

It is obvious from the description above that paradoxes of this type
require not only that tachyons exist, but also that, given an {\em
arbitrary\/} reference frame, it is always possible to send a tachyon
backward in time in that frame.  Obviously, there can be no paradox
if, in one {\em particular\/} reference frame, tachyons can only
propagate forward in time.\footnote{%
Of course such a restriction is anathema in the standard approach
to special relativity since it picks out a preferred frame.
However if you have good physical reasons for picking out a
preferred frame (\eg, the rest frame of the Casimir plates) this
sort of restriction can make good physical sense.}
This situation is exemplified in figure~\ref{fig3}, which shows a
chain of events similar to the one of figure~\ref{fig2}, with the
only difference that now the tachyon has a fixed speed with
respect to the reference frame $\cal K$.  This time, the event
${\cal E}_2$ takes place {\em after\/} the event ${\cal E}_0$ in
$\cal K$. Moreover, since the two events are timelike related,
${\cal E}_2$ follows ${\cal E}_0$ in {\em all\/} frames.  Thus,
there is no paradox, although tachyons are used to send signals.
In the next subsection, we show that the anomalous photons that
give rise to the Scharnhorst effect are precisely ``benign''
tachyons of this type. (See also \cite{konstantinov} for
arguments along similar lines.) More radical proposals to solve
the paradoxes presented by {\em any sort\/} of faster-than-$c$
particles can be found in \cite{recamietal,camenzind} and
references therein. Our goal is much less ambitious, as we
restrict our consideration to Scharnhorst photons only.

\subsection{Scharnhorst photons preserve causality}

In order to show that Scharnhorst photons do not lead to causal
paradoxes, we first introduce an effective metric describing
photon propagation in the Casimir vacuum.  We then show that even
though the light-cones are wider than those of the Minkowski
metric, the spacetime is nevertheless stably causal, which
prevents causal paradoxes from occurring.  If there are multiple
pairs of plates in relative motion, the situation is trickier, and
we analyze it in terms of Hawking's chronology protection
conjecture.

\subsubsection{Effective metric}

The propagation of light signals between perfectly conducting
plates is described by the dispersion relation
$\gamma^{\,\mu\nu}k_\mu k_\nu=0$, where $k_\mu$ is the wave vector
(actually, a one-form), and the coefficients $\gamma^{\,\mu\nu}$
have the form~\cite{oblique}
\begin{equation}%
\gamma^{\,\mu\nu}=\eta^{\mu\nu}+\xi\,n^\mu\,n^\nu\;,%
\label{coeff}%
\end{equation}%
where $\eta^{\mu\nu}$ is the inverse of the Minkowski metric,
$n^\mu$ is the unit spacelike vector orthogonal to the plates, and
$\xi$ is a function. We consider the parallel plates to be
orthogonal to the $x$ axis, so $n^\mu=(0,1,0,0)$.\footnote{
Note the difference with respect to reference \cite{oblique}, where
the plates were taken orthogonal to the $z$ axes.}
The size of the corrections to propagation in the absence of
boundaries can be obtained computing $\xi$ to order $\alpha^2$,
where $\alpha$ is the fine structure constant.  Using the
Euler--Heisenberg Lagrangian one gets~\cite{oblique}
\begin{equation}%
\xi=\frac{11\,\pi^2\,\alpha^2}{4050\,a^4\,m_{\rm e}^4}\approx
4.36\times 10^{-32}\left(\frac{10^{-6}\,{\rm m}}{a}\right)^4\;,%
\label{xii}%
\end{equation}%
where $a$ is the distance between the plates in their reference
frame and $m_{\rm e}$ is the electron mass. Of course, so
tiny a figure means that detecting the effect is a task totally
beyond present-day technology.  However, as we shall now show, the
fact that $\xi$ is positive leads to a speed of propagation
(slightly) larger than $c$, which constitute a potential threat
for causality, no matter how small the deviation is.

The $\gamma^{\,\mu\nu}$ can be interpreted as coefficients of an
effective inverse metric~\cite{oblique,dittrich-gies}. The metric
itself is thus obtained by inverting $\gamma^{\,\mu\nu}$:
\begin{equation}%
\g_{\mu\nu}=\eta_{\mu\nu}-\frac{\xi}{1+\xi}\,n_\mu\,n_\nu\;.%
\label{effmet}%
\end{equation}%
(Warning: We keep raising and lowering indices using the Minkowski
metric $\eta_{\mu\nu}$ and its inverse $\eta^{\mu\nu}$. In
particular, note that $\g_{\mu\nu}\neq\eta_{\mu\rho}\,
\eta_{\nu\sigma}\gamma^{\,\rho\sigma}$.) Since $\xi$ is positive,
the light cone associated to the effective metric $\g_{\mu\nu}$ is
slightly wider, in the direction orthogonal to the plates, than
the one corresponding with $\eta_{\mu\nu}$. This implies that
light travels at a speed $c_\light$ slightly larger than $c$,
except in the case of propagation parallel to the plates.

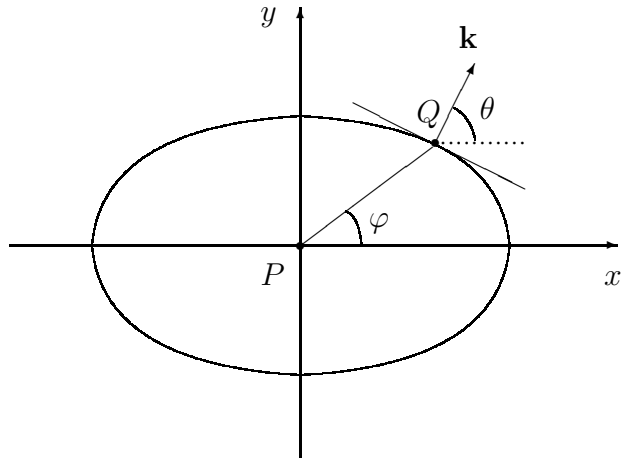
\begin{figure}%
\begin{picture}(150,220)(-30,50)%
\put(70,140){\vector(1,0){230}}%
\put(295,125){$x$}%
\put(180,60){\vector(0,1){170}}%
\put(165,225){$y$}%
\put(180,140){\circle*{3}}%
\put(165,125){$P$}%
\bezier{500}(101,140)(105,185)(180,189)
\bezier{500}(180,189)(255,185)(259,140)
\bezier{500}(101,140)(105,95)(180,91)
\bezier{500}(180,91)(255,95)(259,140)
\put(200,194){\line(2,-1){65}}%
\put(231,179){\circle*{3}}%
\put(224,187){$Q$}%
\put(180,140){\line(4,3){50}}%
\put(231,179){\vector(1,2){15}}%
\put(240,215){$\bf k$}%
\put(231,179){\circle*{1}}%
\put(234,179){\circle*{1}}%
\put(237,179){\circle*{1}}%
\put(240,179){\circle*{1}}%
\put(243,179){\circle*{1}}%
\put(246,179){\circle*{1}}%
\put(249,179){\circle*{1}}%
\put(252,179){\circle*{1}}%
\put(255,179){\circle*{1}}%
\put(258,179){\circle*{1}}%
\put(261,179){\circle*{1}}%
\put(264,179){\circle*{1}}%
\bezier{100}(197,153)(202,151)(203,140)%
\put(206,147){$\varphi$}%
\bezier{100}(238,192)(244,189)(246,180)%
\put(248,188){$\theta$}%
\end{picture}%
\caption{\small Wavefront (or line of constant phase) for a light
signal emitted at the point $P$.  Note that the wave vector ${\bf k}$
forms a non-vanishing angle $\theta-\varphi$ with the direction of
propagation.}%
\label{fig4}%
\end{figure}

The exact value of this speed in the rest frame of the plates can
be obtained by considering the way a light signal is actually sent
from a point $P$ in space to another point $Q$.  A pulse at $P$
originates a wavefront that propagates outwards and eventually
reaches $Q$.  The speed of light in the direction $PQ$ is then
given by the ratio
\[\frac{\mbox{distance between $P$ and $Q$}}{\mbox{time taken by
the wavefront to reach $Q$}}\;.\]%
(Of course, the speed of light is independent of position, because
$g_{\mu\nu}$ turns out to have constant coefficients.)  Assuming
that the pulse starts at $t=0$ at the point $P$ with coordinates
$(0,0,0)$, the wavefront is described by the equation
\begin{equation}%
S(t,{\bf x})=-ct+\left(g_{ij}\,x^i x^j\right)^{1/2}=0\;,%
\label{eikonal}\end{equation}%
where Latin indices run from 1 to 3, and the function $S$ --- the
eikonal --- is a solution of
\begin{equation}%
\gamma^{\mu\nu}\;\partial_\mu S\;\partial_\nu S=0\;.%
\label{eqforeikonal}\end{equation}%
Thus, the time required for propagation from $P$ to the point $Q$
with coordinates $(x,y,z)$ is
\begin{equation}%
ct=\left(g_{ij}\,x^i x^j\right)^{1/2}={\bf
x}^2-\frac{\xi}{1+\xi}\,\left({\bf n}\cdot{\bf x}\right)^2
=\left|{\bf
x}\right|\left(1-\frac{\xi}{1+\xi}\,\cos^2\varphi\right)^{1/2}\;,
\label{time}\end{equation}%
where $\varphi$ is the angle that $PQ$ forms with the normal to the
plates, ${\bf n}$.  The speed of propagation for light in the
direction $PQ$ is then
\begin{equation}%
c_\light(\varphi)=\frac{\left|{\bf
x}\right|}{t}=c\left(\frac{1+\xi}{1+\xi\,\sin^2\varphi}\right)^{1/2}\;.
\label{cl}\end{equation}%
Note that $c_\light(\varphi)$ is different from the phase velocity
\cite{oblique}
\begin{equation}%
v_{\rm phase}(\varphi)=c\left(1+\xi\,\cos^2\varphi\right)^{1/2}
\label{phasevelocity}\end{equation}%
corresponding to an angle $\varphi$ between the wave vector ${\bf
k}$ and ${\bf n}$.  This discrepancy is due to the fact that the
surfaces of constant phase are ellipsoidal rather than spherical,
which leads to a tilt between the direction of propagation $PQ$
and the wave vector ${\bf k}=\mbox{\boldmath $\nabla$}S$ (see
figure~\ref{fig4}).

The Scharnhorst photons travel at a definite speed in any given
reference frame.  This is a trivial consequence of the previous
result, that they have a well-defined speed $c_\light(\varphi)$ in
the reference frame at rest with respect to the plates, and of
the relativistic law for the composition of velocities. However,
one can check this explicitly using the description of photon
propagation given by the effective metric (\ref{effmet}).

For two events with coordinates $\{x^\mu\}$ and $\{x^\mu+\delta
x^\mu\}$, lying along the worldline of a photon, we have
$\g_{\mu\nu}\delta x^\mu\delta x^\nu=0$. An observer with
four-velocity $u^\mu$ will decompose the spacetime displacement
$\delta x^\mu$ into a time lapse
\begin{equation}%
\delta\tau=-\frac{1}{c}\,\eta_{\mu\nu}u^\mu\delta x^\nu%
\end{equation}%
and a space displacement
\begin{equation}%
\delta\chi=\left(\eta_{\mu\nu}\delta x^\mu\delta
x^\nu+c^2\,\delta\tau^2\right)^{1/2}\;,%
\end{equation}%
respectively.  The photon speed $c_\light^{(u)}$ with respect to
the observer $u^\mu$ is given by
\begin{equation}%
c_\light^{(u)} =
\lim_{\delta\tau\to0}\frac{\delta\chi}{\delta\tau} =
c\,\lim_{\delta\tau\to0}\left[1+\frac{\xi}{1+\xi}
\left(\frac{n_\mu\delta x^\mu}{u_\nu\delta
x^\nu}\right)^2\right]^{1/2}\;,%
\label{speed}%
\end{equation}%
where the relationship
\begin{equation}%
\eta_{\mu\nu}\delta x^\mu\delta
x^\nu=\frac{\xi}{1+\xi}\,\left(n_\mu\delta x^\mu\right)^2\;,%
\end{equation}
which follows from (\ref{effmet}), has been used.  The important
points about equation~(\ref{speed}) are that $c_\light^{(u)}\geq
c$ always, and that $c_\light^{(u)}$ has only one value for each
observer $u^\mu$.  The latter feature prevents the possibility of
causal paradoxes, which require signals that travel with the {\em
same\/} speed greater than $c$ in two different reference frames.

\subsubsection{Stable causality}

Adopting the formalism of the effective metric (\ref{effmet})
allows us to borrow many of the notions of general relativity,
and to conclude in a more straightforward way that Scharnhorst
photons cannot violate causality. In particular, since in general
relativity the global causal structure is not specified by the
field equations, considerable thought has gone into how to
characterize causality~\cite{Hell}. We only need one key
definition, and one key textbook result.

\paragraph{Definition: Stable causality.}

A spacetime is said to be stably causal if and only if it
possesses a Lorentzian metric ${\rm g}_{\mu\nu}$ and a globally
defined scalar function $t$ such that $\nabla_{\!\mu} t$ is
everywhere non-zero and timelike with respect to ${\rm
g}_{\mu\nu}$. (See reference~\cite{Hell}, p.\ 198.)

\paragraph{Theorem:}

A stably causal spacetime possesses no closed timelike curves, and
no closed null curves.  (See reference~\cite{Hell}, p.\ 199.)

\bigskip
The relevance of this definition and theorem is that spacetime,
endowed with the Minkowski metric (\ref{mink}) outside the plates,
and with the effective metric (\ref{effmet}) inside the plates, is
stably causal. Indeed, using the time coordinate in the rest frame
of the plates as the globally defined $t$, then $\nabla_{\!\mu} t$
is everywhere non-zero and timelike. Therefore causality is safe.

This is enough to guarantee that you can never encounter a
causality violation if you only have a single pair of Casimir
plates to deal with, or even multiple plate pairs that do not
intersect and do not move with respect to each other. Multiple
pairs of Casimir plates in relative motion will be a little
trickier. Far in the past, when the pairs are well separated,
there is certainly no causality violation and the question arises
as to what happens as the various pairs of plates approach each
other.

\subsubsection{A paradox with multiple pairs of moving plates?}

Suppose we have {\emph{two}} pairs of Casimir plates, moving with
relative speed $v_\relative$: Let the usual speed of light be $c$
and the speed of the Scharnhorst photons be $c_\light > c$. Then
the usual textbook exercises presented in elementary courses of
special relativity would seem to indicate the {\emph{risk}} of
causality violations whenever~\footnote{%
This condition on the velocities simply comes from the Lorentz
transformation of the time coordinate
\[
t \to t' = \gamma \; ( t - v_\relative x/c^2).
\]
Setting $x = c_\light \; t$, and asking for $t' < 0$, implies
$v_\relative \; c_\light > c^2$. One then applies the same logic
to the return trip to find a necessary condition for the
existence of closed timelike curves.}
\begin{equation}
v_\relative \; c_\light > c^2. \label{E:paradox}
\end{equation}
But before we deduce the existence of an actual logical paradox we
should look more carefully at the spacetime diagram.

Suppose the plates are truly infinite in extent, as they should be
for the Casimir vacuum to be exact. Then to generate the causal
paradox of figure \ref{fig2} the entire closed timelike loop $\E_0
\to \E_1 \to \E_2 \to \E_0$ must lie inside {\emph{both}} pairs of
Casimir plates. That is, the two Scharnhorst vacua must be
inter-penetrating, in particular the conducting plates required to
set up the Casimir vacuum must be able to pass through each other
without affecting each other. But in that case, the symmetries
leading to the effective metric in equation (\ref{effmet}) have
been violated, so there is no reason to believe that the effective
metric will look anything like (\ref{effmet}), and there is no
reason to believe that any closed timelike loop forms.\footnote{%
In particular the region containing the putative closed timelike loop
$\E_0 \to \E_1 \to \E_2 \to \E_0$ is simultaneously subject to two
different and incompatible effective metrics, so the logic that at
first glance seems to lead to closed timelike loops is actually
internally inconsistent.}

We could try to be a little more realistic by taking two pairs of
half-infinite plates. To be specific let one pair of half-infinite
plates be confined to the region $x>0$, while the second pair is
confined to the region $x<0$. Then one could try to set up a causal
paradox by, for instance, sending messages to the left using the
vacuum corresponding to the $x>0$ plates, and sending messages to the
right using the $x<0$ plates. (To completely close the loop we would
also need to send two messages along the $x$ axis, to get from the
$x>0$ vacuum to the $x<0$ vacuum and back.)

The problem in this case is edge effects: the effective metric
given in (\ref{effmet}) is only expected to be a good
approximation far away from the edge of the plates, while well
outside the plates the effective metric should approach that of
Minkowski space. Near the edge of the plates the effective metric
is impossible to calculate, and the situation only gets worse when
two pairs of half-infinite plates pass each other with a grazing
not-quite collision. It is certainly clear that the simple naive
result of equation (\ref{E:paradox}) should not be trusted.

The present arguments do not guarantee the total absence of
causality violations, but they do demonstrate that the most naive
estimates of the causality violating regime are likely to be
grossly misleading.

\subsubsection{Chronology protection}

To actually demonstrate that causality violation is excluded, at
least if we stay within the realm of semiclassical quantum field
theory (and this is the underlying approximation made to even set
up the entire formalism), we invoke the notion of ``chronology
protection'' in the sense of Hawking~\cite{CPC-Hawking,CPC-Wald,%
CPC-Sushkov,CPC-Visser}.

The point is that when the two pairs of plates are well separated,
long before the two pairs approach each other, they are individually
stably causal and there is no risk of closed timelike curves --- the
``initial conditions'' are that the universe starts out with sensible
causality conditions. So if a region of closed timelike curves forms
as the two pairs of plates approach each other, that region must have
a boundary, and there must be a ``first'' closed null curve.

Hawking has argued that the appearance of the first closed null curve
will lead to uncontrollable singularities in the renormalized quantum
stress-energy as vacuum fluctuations pile up on top of each
other~\cite{CPC-Hawking}. (Roughly speaking, because of the existence
of a closed null curve, a single photon [real or virtual] could meet
up with itself and coherently reinforce itself an infinite number of
times.) Though Hawking's argument was presented in the context of the
causal problems typically expected to arise in Lorentzian wormholes,
the argument is in fact generic to any type of ``chronology horizon''.

A second, slightly different, version of chronology protection
has been formulated by Kay, Radzikowski, and
Wald~\cite{CPC-Wald}. Technically, they demonstrated that at the
chronology horizon the two-point function (Green function) fails
to be of Hadamard form. As a consequence they could argue that the
renormalized stress tensor does not even make sense at the
chronology horizon, and that the entire formalism of quantum field
theory on a fixed background spacetime suffers total failure.
(Note that in this approach the stress-energy tensor may remain
bounded in the vicinity of the chronology horizon, though it is
guaranteed to be ill-defined at the chronology horizon itself. See
also~\cite{CPC-Sushkov}.)

One of the present authors has further argued that in the vicinity of
any chronology horizon there is an invariantly defined ``reliability
horizon'' beyond which Planck scale physics comes into
play~\cite{CPC-Visser} --- the region characterized by the existence
of closed spacelike geodesics shorter than one Planck length is a
region subject to violent Planck-scale fluctuations, even if the
expectation value of the stress-energy is small.

All these variants on the idea of chronology protection agree on
one thing: If the geometry is to remain well-defined so that
semiclassical physics makes sense, which is certainly necessary to
even define the basic precursors leading to the notion of
causality, then the chronology horizon must fail to form.  Indeed,
if the chronology horizon somehow does manage to form, then
causality paradoxes are the least of your worries since you are
automatically driven into a regime where Planck-scale quantum
gravity holds sway.

\section{Conclusion}
\label{S:conclusion}
\setcounter{equation}{0}

In this paper, we have presented a critical assessment regarding
the possibility, from the point of view of the basic physical
principles of relativity and causality, of faster-than-$c$
signalling. Our analysis is motivated by some recent theoretical
predictions of ``superluminal'' photon propagation \cite{scharnhorst,%
barton-scharnhorst,drummond-hathrell}.  We have shown that such
effects are kinematically compatible with special relativity,
because the latter requires only the existence of an invariant
speed, not necessarily a maximum one. Also, they do not lead to
causal paradoxes, which can arise only from tachyons whose speed
has no fixed value in a given reference frame. Perhaps
surprisingly, it is the soft breaking of Lorentz invariance, which
always occurs in these effects, that fixes the value of the speed
of the faster-than-$c$ photons in a specific frame and thus saves
causality.  Such a breaking is, however, due to the choice of a
vacuum state and does not extend to the invariance group of the
fundamental physical laws, which still remains the usual $C=c$
Lorentz group.

\appendix
\section*{Appendix A: Alternative kinematical groups?}

The derivation of Lorentz transformations {\em \`a la\/} von
Ignatowsky, on which our analysis in section~\ref{Ss:ignat} relies,
shows how robust the structure of the Lorentz group is. In order to
obtain different transformations, at least one of the hypotheses
(i)--(iii) listed at the beginning of section~\ref{Ss:ignat} must fail
to apply. This remark is relevant in the context of the now
fashionable attempts at replacing the Lorentz group by a different set
of transformations, in order to provide a kinematical basis for a new
physics that is not Lorentz-invariant at high energies~\cite{ac},
which might be needed to account for the observed overabundance of
cosmic-ray TeV photons and the missing GZK cut-off in a natural way.
(See \eg, \cite{Sigl} and relevant references therein.)  The hypothesis
that is most likely to fail is (ii) --- the relativity principle.
However, since (ii) enters von Ignatowsky's argument merely as the
requirement that the set of transformations between inertial frames
form a group, it is obvious --- indeed, tautological --- that no group
structure can emerge if one rejects it.  In fact, it is a result of
the analysis performed in references~\cite{ignat1,fr1,terletskii,%
sussmann,berzi,zecca,leekalotas,ll,rindler,jammer,torretti} that the
Lorentz and Galilei transformations are the {\em only\/} ones that
satisfy (i)--(iii).

We must add a caveat, though. In addition to (i)---(iii) there are
other, more fundamental, hypotheses underlying {\em any\/}
derivation of the Lorentz transformations.  These are the (usually
implicit) assumptions that measurements of time intervals are
expressed by arbitrarily small real numbers, that it is possible
to describe space in terms of Cartesian frames, and that frames
can move with respect to each other at an arbitrary speed $v<C$.
The first two assumptions require that there is no fundamental
scale in Nature. The third one is a consequence of combining scale
invariance with the relativity principle. However, if there were a
fundamental length scale in the physical universe, say $l_0$,
under which no measurement could be performed, it would be
meaningless to consider Lorentz transformations at speeds
arbitrarily close to $C$.  Indeed, consider two frames $\cal K$
and ${\cal K}'$ which define a length unit following the same
procedure --- a hypothesis implicit in special relativistic
kinematics, as we discussed in section~\ref{physcoord}.  It is
always possible to find a relative speed which is so large that
the unit of ${\cal K}'$ as measured from $\cal K$ turns out to be
smaller than $l_0$, in which case no comparison between
measurements can be performed.  Thus, it might be that, although
hypotheses (i)--(iii) are still satisfied, von Ignatowsky's proof
of the necessity of the Lorentz group is nevertheless vitiated by
rejecting some assumption which is even more basic.  This leaves
open the possibility that non-Lorentzian kinematical groups could
somehow be developed.

\section*{Appendix B: Maximum speed and negative energy densities}

Although not crucial for the consistency of special relativity, it
would nevertheless be important to understand whether there are
physical bounds for the speed of signals.  The possibility of
superluminal effects is usually associated with special conditions
on the expectation value of the stress-energy-momentum tensor for
the quantum fields in the modified vacuum state.  Indeed, in
general homogeneous non-trivial vacua, the inverse effective
metric for photon propagation takes (at lowest nontrivial order)
the form
\[
\gamma^{\mu\nu}=\eta^{\mu\nu}-Q\langle T^{\mu\nu}\rangle\,,%
\]
where $Q$ is, in general, a function of the variables and parameters
of the Lagrangian~\cite{dittrich-gies}.  In the particular case of the
Scharnhorst effect $Q$ turns out to be (to order $\alpha^2$) a
positive constant.  It is then clear from the dispersion relation
$\gamma^{\mu\nu}k_\mu k_\nu=0$, together with the above, that in order
for the modified light cones to be wider than those of
$\eta_{\mu\nu}$, one must have $\langle T^{\mu\nu}\rangle
k_{\mu}k_{\nu}<0$. Since $k^{\mu}$ is a null vector with respect to
the Minkowski metric, it follows that the Scharnhorst effect implies a
violation of the null energy condition.

If this circumstance were generic, \ie, if the increase of the
speed of light were always associated with the presence of
negative energy densities, due to vacuum
polarization~\cite{latorre,dittrich-gies}, a possible maximum
velocity should correspond to a maximum negative energy density.
The problem becomes then whether there is a lower bound for the
energy density. One can envisage several possibilities that
would lead to this conclusion: quantum field theory itself could
break down when the vacuum energy density is too high (in absolute
value), or something like the Ford--Roman quantum
inequalities~\cite{Ford-Roman} could apply.\footnote{
If the amount of energy which can be confined in a given spacetime
volume $V^{(4)}$ is limited by the uncertainty principle, one
would dimensionally expect a relationship of the type
$\langle\rho\rangle > -\hbar/V^{(4)}$.}
It is also conceivable that non standard geometrical terms
may appear in higher order corrections to the Casimir effect
and lead, at ultra-short separations of the plates, to positive
energy densities. Of course, in this latter case the maximum speed
would be a geometry- (as well as vacuum-) dependent quantity,
rather than a fixed constant of nature.  Any of these
possibilities, or any other one giving a lower bound on the energy
density, would in turn lead to an upper bound for the photon
speed.

\section*{Appendix C: Reprise --- What is $C$?}

Between Casimir plates, the internal constitution of standards of
time and distance based on the electromagnetic interaction is
certainly influenced by the same processes that lead to the
Scharnhorst effect.  This might turn into an ``invisibility'' of
the effect itself, similarly to what happens, according to
Lorentz, in a reference frame that moves with respect to the
\aether, when dynamical contraction of rods and slowing of clocks
do not allow one to measure the ``true'' speed of
light~\cite{lorentz}. In other words, measurements performed using an
equipment based on electromagnetism might lead to the empirical result
that light travels always at the speed $c=2.99792458\times
10^8\,\mbox{m s}^{-1}$, even though our theories imply that the speed
is $c_{\rm light}$, not necessarily identical to $c$. However, one can
use units which are unaffected by the presence of plates --- for
example, based on gravitational physics --- which would still measure
$c$ as the invariant speed, thus making possible to detect the
Scharnhorst effect, at least in principle.

Nevertheless, the possibility that it is $c_\light$, rather than
$c$, the velocity that would turn out to be invariant by
measurements based on electromagnetic standards, seems to suggest
that, as far as such measurements are concerned, one should
consider a Lorentz group with $C=c_\light$. However, since
$c_\light\neq c_\gravity$, this Lorentz group would not be the one
appropriate to define local Lorentz invariance in general
relativity.  (Of course, the same situation would occur if photons
and gravitons were travelling at different speeds.)  Furthermore,
there would be two possible candidates to the role of invariant
speed $C$ in {\em special\/} relativity.  {From} the arguments in
section~\ref{physcoord} it follows that one cannot choose {\em a
priori\/} whether $C=c_\light$ or $C=c_\gravity$, because the
actual value of $C$ is related to the behavior of real clocks and
rods.  Likely, electromagnetic clocks/rods will correspond to
$C=c_\light$, while gravitational clocks/rods should lead to
$C=c_\gravity$. Since it seems hard to live with {\em two\/}
relativity theories, the most natural step to take in this case
would be perhaps to reconsider the role of Lorentz invariance,
regarding it as a symmetry property of specific theories rather
than as a fundamental meta-principle of all physics.

\section*{Acknowledgements}
The research of Matt Visser was supported by the US Department of
Energy. Stefano Liberati was supported by the US National Science
Foundation.  SL wishes particularly to thank M.~Testa and T.~Jacobson
for stimulating discussions.

\end{document}